\documentstyle[preprint,aps]{revtex}
\begin{document}
\draft
\title{
    Universal Elementary Constituents of Potential Transformations Shifting
Waves\\
       (qualitative Theory) }
\author{
    V. M. Chabanov, B. N. Zakhariev}
\address{ Laboratory of the Theoretical Physics,
 Joint Institute for Nuclear Research  Dubna, 141980, Russia;
e-mail: zakharev@thsun1.jinr.dubna.su ; URL: 
$http://thsun1.jinr.ru/\sim zakharev/$} \author{ S. A. Sofianos} \address{ 
Department of Physics, University of South Africa, Pretoria, South Africa} 
\date{}
\maketitle
\begin{abstract}
It is shown that the potential perturbation that shifts a chosen standing
wave in space is a block of potential barrier and well for every wave
bump between neighbouring knots. The algorithms
shifting the range of the primary localization of a chosen bound state
in a potential well of finite width are as well applicable to the
scattering functions  if  states of the  continuous spectrum are
considered as bound states normalized to unity but distributed on
an infinite interval with vanishing density. The potential perturbations of
the same type  on the half-axis concentrate the scattering wave at the
origin, thus creating a bound state embedded into the  continuous
spectrum (zero width resonance). It is an improved version of paper: 
Annalen der Physik, 6 (1997) 136. 
\\\\

\end{abstract}

\section{Introduction}
In the recent years it became clear \cite{Zak1,Zak2,Zak5} what qualitative
transformations of any initial potential $V(x)$ are required
to  change an energy level or to shift the region
of  main concentration of any  bound state in configurational space
(i.e. the distribution of the corresponding standing wave over the $x$-axis,
cf. for example Fig.1a,b).
This means in effect the extraction of physical essence from
the mathematical apparatus of the quantum inverse problem
(Gelfand-Levitan-Marchenko formalism)\cite{Newton,Zak1,Saba}, Darboux transformation,
factorization method and supersymmetry theory
\cite{Cooper,Andri,Baye,Sukumar,Po}.
It was done via visualization
of analytical formulae corresponding to exactly solvable models.
The simple and universal  rules of
qualitative prediction of the transformed potentials and resulting systems
systems (intuitive quantum design \cite{Zak1,Zak2,Zak5}) go beyond
the scope of these models.

It is also possible to control the features of states of the continuous
spectrum (reflection, transparency, resonance parameters \cite{Zak2}).
In the present paper it is shown that  an additional
potential $\Delta V(x)$  shifting the  $n$--th bound
state over $x$ axis to the right (left) can be considered as a sum of
$n$ elementary
universal blocks consisting of a potential barrier  and a well. We show below
that perturbations constructed out of these blocks  can  "gather" also
the scattering states rolled out on the half-axis
confining them into bound states embedded  in the continuous
spectrum (BSEC) at a fixed energy; these states can be considered as 
resonance states with vanishing width.

  We discuss below the qualitative aspects of the theory in
Sec. II, give the corresponding formalism
  in Sec. III and  summarize  our conclusions in Sec. IV.

\section{Qualitative considerations}
\subsection{Shifting of standing waves}

 We have found that the perturbation of a potential shifting the
ground state localisation
in space (see Fig. 1c,b for the case of initial infinite rectangular well)
consisting of a well and a barrier can be considered as an universal
elementary block for shifting the localisation of arbitrary bound states
over $x$ axis  without actually shifting any
energy level (here the influences of barriers and wells are mutually
compensated).
To shift, for instance,  the second bound state to the left we need two
such blocks (Fig. 1f). The
n-th bound state can be shifted with n blocks.  Every block corresponds to
one bump in the initial bound state standing wave. The boundaries of each
block coincide with the relevant neighboring knots of the wave function.
The amplitudes of the potential oscillations are decreasing together with
the transformation of the standing wave which is concentrating  towards the
origin. The same situation holds for shifting waves to the right, but with
mirror reflection of potentials and wave functions: barrier-well block for 
each wave bump instead of wave-barrier blocks.\\

   Now we can see that potentials confining a BSEC
are also constructed of an infinite number of these universal blocks
with slowly decreasing amplitudes as $\sim 1/r$ (see Fig. 2).
  Neuman and Wigner \cite{Neu} were the first who obtained formulae
for BSEC. Further work on BSEC was   undertaken by a number of other authors
(see [11-13] and references therein and experimental paper [15]). We conclude
 that the construction of BSEC can be considered as
"gathering" scattering states and concentrating them in the vicinity of the
origin.

\subsection{Scattering functions as "bound" states }
Let us consider the most simple case of a continuous spectrum wave function
for a free particle. The function
describing free motion on the half--axis $0<r<\infty $ at fixed
energy $E_{b}=k_{b}^{2}$ is given by  $\sin(k_{b}r)/k_{b}$.
Usually the scattering functions are normalized to a $\delta $--function.
However, we shall show that one can treat them as bound states
normalized to unity. \\

If an infinite unpenetrable potential wall would be placed at the point of
the first
knot $r=\pi /k_{b}$ of this sine--function, then the first half--wave of
this sine will coincide with the bound ground state $\psi (E_{b},x)$ formed
by an infinite potential well. In order to normalize
the obtained function to unity one simply  multiplies the sine by
$\sqrt {2k_{b}/ \pi }$. Let us now move the potential wall to the point
$r=2\pi/k_{b}$ of the second knot of the free wave.
Now on the finite segment of double length the sinusoid $\sin(k_{b}r)$
will coincide exactly with the wave function of the second bound state
in a more wide well. The  required normalization is then given by
$\sqrt{k_{b}/ \pi }$. If we continue to expand the well we will
get, at the $n$th step, the $n$th bound state of the rectangular well
coinciding with a chosen sinusoid and the corresponding normalization factor
will be $\sqrt {2k_{b}/(n \pi )}$. In the limit $n\to \infty $ we will
obtain a "bound" state $\psi (E_{b},x)$ of an infinitely wide rectangular well
with "vanishing" normalization factor while normalized to unity,
\begin{equation}
         \int_{0}^{\infty}\psi ^{2}(E_{b},x)dx=1 \,.
\end{equation}
This bound state is distributed over the infinite interval with
vanishing density.
The situation is somewhat analogous to the
infinitely narrow but normalized to unity peak of a $\delta $--function
\begin{equation}
         \int_{-\infty}^{\infty}\delta (x)dx=1\,.
\end{equation}
  Using inverse problem or SUSY algorithms  \cite{Zak1,Po,Baye,Stahl,Still}
\cite{Andri} 1991,  each of the bound states
considered above could be moved to the left wall of the infinitely deep well
using blocks of transformations of the potential described above.
The derivative  of a normalized wave
function at the origin can be done arbitrarily large. Fig. 2a exhibits
the functions of BSEC by consecutive growth of $\psi '(E_{b},0)$ and
corresponding perturbations of the potential. For a higher concentration of the
function near the origin perturbations $V(x)$ with increasing
amplitudes of oscillations are needed, see Fig. 2b.\\

  This view allows to bridge the differences in the notions of bound
and continuum states.

  An interesting fact is that the knots of the BSEC do not move by 
increasing $\psi '(E_{b},0)$, a phenomenon which must have a clear 
 qualitative explanation.

  The blocks of  potentials producing BSEC's  can be arbitrarily rearranged
without loosing the gathering property. Then the wave functions
on the corresponding intervals must also be rearranged. We must only change
their amplitudes for smooth continuation through the junction points (block
boundaries). Some of the blocks can also be repeated; an interesting limiting
case is the periodic potential on the half-axis with the bound state at the
boundary of the conduction band. Some blocks can be  missed, i.e. substituted
by intervals with zero potential. The tail of the gathering potential
can be deleted from some point. Then BSEC will transform into a
quasi-bound (resonance) state with finite width which must be smaller the
further the tail is cut. The system can be continued to the whole
axis using  mirror reflections of blocks and corresponding solutions to
create BSEC decreasing in both opposite directions $x\to \pm \infty $.\\

  Scattering wave functions  at  energies neighboring to those of a BSEC
become asymptotically ($x\to \infty $) free waves without
phase shift. This effect  can be explained  by total reciprocal
compensation of the influence of intervals where the wave is gathered
and intervals from which the wave is forced out
 (see beatings in wave function shown in Fig. 3).
The period of such beatings increases without bound when approaching the 
energy and in the limit there
$k_{b}^{2}$  of the BSEC, $lim \psi (r,k \to k_{b})= \sin(k_{b}r)/k_{b}$.
So, in the limit there is left only one beating on the full half-axis that
is what presents BSEC (Fig. 2).

\subsection{Background Formulae}
  It is the purpose of this paper to clarify the
general qualitative peculiarities of evolution of potentials and wave
functions by transformations shifting waves at fixed energies which where not
mentioned before.
We thus give here only the final formulae only (without any derivation
which is given, for example, in~\cite{Zak1,Zak3}).
The transformation of the  potential  caused by  the change of the
spectral weight parameter from $c_{0,\nu }$ to $c_{\nu }$ ($c_{\nu }=
\psi '(E_{0,\nu},0)$ where $\psi (E_{0,\nu },x)$ is the normalized
wave function at the chosen energy) is given by
\begin{equation}
    V(r)=  V_0(r) - 4\delta c_\nu^2\, \varphi_0'(E_{0_,\nu},r)
      \varphi_0(E_{0,\nu},r)\,p^{-1}(r)+
      2(\delta c_\nu^2)^2\,
     \varphi_0^4(E_{0,\nu },r)\,p^{-2}(r)\,,
\label{v}
\end{equation}
where
\begin{equation}
    \delta c_{\nu }^{2}=c_{\nu }^{2}-c_{0,\nu}^{2};\,\,\,\, c_{\nu}^{-2}
=\int_{0}^{\infty} \phi^{2}(E_{0,\nu },r) dr \end{equation}
and
\begin{equation}
     p(r)=1 +  \delta c_{\nu }^{2}\int_0^r \varphi_0^2(E_{0,\nu},t)dt\,.
\end{equation}
The corresponding perturbed regular solution  fulfilling the boundary
conditions
$\varphi (E,0)=0$ and $\varphi '(E,0)=1$  at an  arbitrary $E\ne E_{0,\nu}$,
is given by
\begin{equation}
    \varphi (E,r)  = \varphi_0(E,r) -\delta c_\nu^2\,
    \varphi_0(E_{0,\nu },r)\,p^{-1}(r) \int_0^r
    \varphi_0(E_{0,\nu },t) \varphi_0(E,t)dt\,.
\label{phi}
\end{equation}
The regular solution at $E_{0,\nu}$ is
\begin{equation}
    \varphi(E_{0,\nu},r) =\varphi_0(E_{0,\nu},r)\,p^{-1}(r)\,.
\label{phin}
\end{equation}
The BSEC for initial repulsive Coulomb potential is shown in Fig.5. 
Shifting of bound  states  in atrective Coulomb potentials was considered 
in \cite{Fer}. The case of a creation BSEC on a slope of the initial linear 
potential had been considered in the paper \cite{Calo}. For this case 
the potential perturbations confining the bound state  grow without limit 
by an amplitude which moves away from the  direction  of the turning point 
of the barrier.\\

It is interesting to consider the multichannel systems and find there
also
some universal blocks of potential transformations. It is easy to generalize
one-channel case to the multi-channel one with equal channels. In the
general case it is an open problem.
\section{conclusion}
We have shown that the additional potential $\Delta V(x)$ which shifts
$n$--th bound state consists of $n$ elementary blocks (barrier +
well), one of which is shown in Fig. 1c, shifting the ground state.

We have managed to extract the physical essence from proper formulas,
reveal universal  "atoms" of potential perturbations, performing such
transformations (quantum "design"). It is remarkable that then distinct
islands of quantum intuition began gradually merging into united qualitative
theory of spectral, scattering and decay control.
  The BSEC's are explained as scattering states normalized to unity and
gathered to the origin.
  The zero scattering phase shift for solutions with energies near to BSEC
are simply explained as result
a total reciprocal compensation of gathering and pulling
under the potential oscillations.\\

It is of interest to find some universal building blocks   in
multi--channel
systems. In this case the  different and coincident thresholds or
closed channels could be considered. The first results in this direction
were already achieved \cite{Zak2,Stroh}.


\acknowledgements
The authors gratefully  acknowledge financial support from the
Russian Funds of Fundamental Research (RFFR),
Soros Fund, the Foundation of Research and
Development, and the University of South Africa.
\newpage
\begin{center}
                 FIGURE CAPTIONS
\end{center}

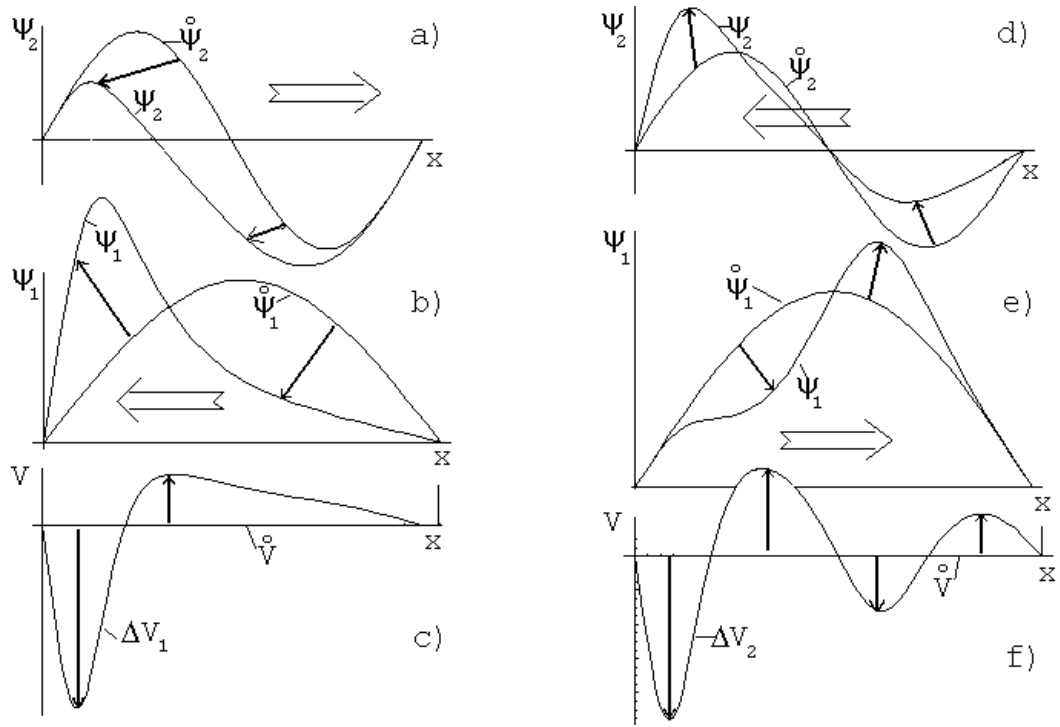
Fig. 1:
     Increase in the spectral weight parameter (the derivative 
$\Psi_{i}'(x=0)$) of a chosen bound state related to initial rectangular 
potential well with conservation of the normalization of the wave function 
leads to the concentration of distribution of probability for given state 
in a near vicinity of the origin. b) and d) -- the transformation of the 
chosen ground and the second states  c) and f) -- corresponding 
perturbations $\Delta V_{1,2}(x)$ of the flat bottom of the initial well; 
a) and e) -- examples  of a reaction of all states except for the chosen 
one:  second and the first states $\Psi _{2,1}$ which undergo the recoil in 
opposite direction in comparison with the chosen one. The thick arrows ==> 
and <==  show the direction of the average localization shifts of the 
states.  In the limit $\Psi '(x=0)\to \infty $ there occurs "pressing" of a 
chosen state into the wall at the origin.

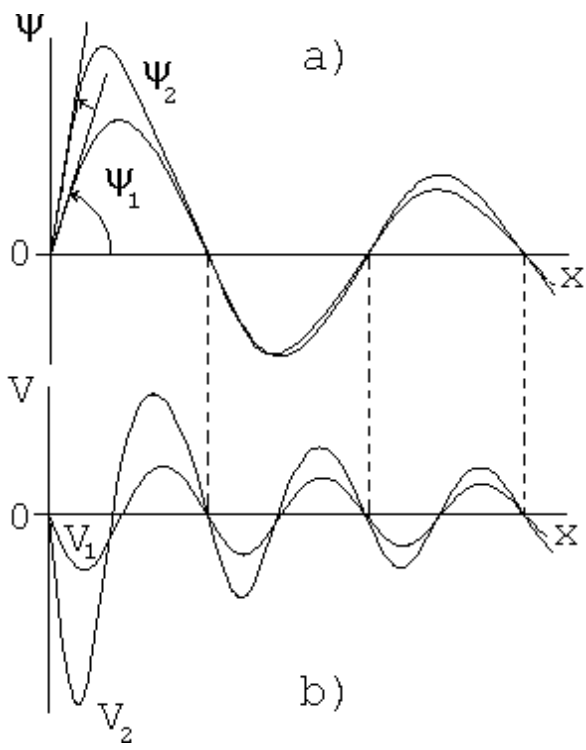
Fig. 2:
      The potentials b) performing a confinement of the bound states
in continuous spectrum at the fixed energy $E_{b}$
     and the proper wave functions a)
    for  different values the weight spectral factors
"c" (values of derivatives at the origin).
With growth of "c" BSEC are more and more pressed to the origin.

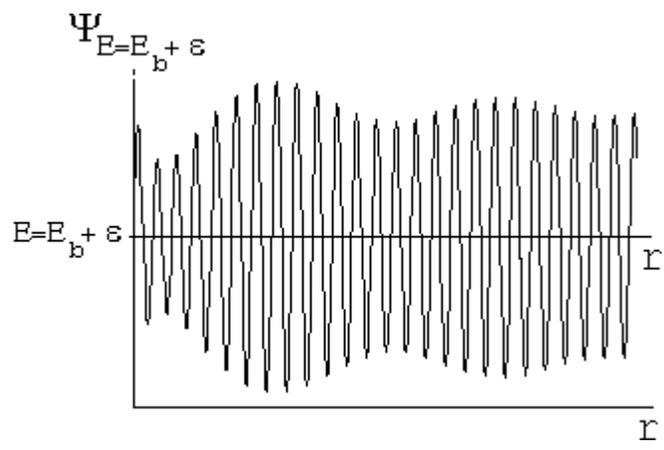
Fig. 3:
  Scattering function related to  energy $E_{b}+\varepsilon $ that is
close to BSEC energy $E_{b}$. The correlation of their oscillations
with potential ones is violated,
      the properties of the potential to gather wave function remain
 only on finite sections of the half-line $r$
      which leads to "gathering"
them but on finite parts of co-ordinate axis.
Out of these sections the oscillations of the functions and the potential
 appear to be opposite in phase and there occurs ejecting the wave
function to nearest  regions of wave gathering.
This develops in beatings dying
at infinity with decreasing of potential. The asymptotic of such functions
coincides with unperturbed solutions (with zero phase shifts) due to mutual
compensation of "concentrations" and "displacements".

Fig. 4

      (a) A free motion wave function  $\sin(k_{b}r)$ at
fixed energy in continuous spectrum $E_{b}=k_{b}^{2}$.

      (b) An infinite unpenetrable potential wall at the point
of the first knot of $\sin(k_{b}r)$ creates infinitely deep
potential well in which
scattering half-wave coincides with  the wave function of bound sate.

       (c) When shifting the wall in $n$th knot, $\sin(k_{b}r)$ is
already $n$th bound state of the new well and strictly coincides
with the free scattering solution inside the well. In the limit
$n \to \infty $ the scattering function can be considered as a
bound state, distributed on the whole half-axis $0<r<\infty $,
only if one needs to normalize it by unity, vanishing
normalization factor $\sim 1/\sqrt{n}$ should be inserted.

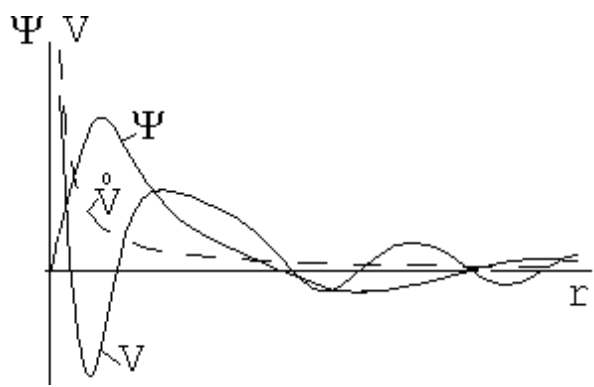
Fig. 5:
   Transformation of initial repulsive coulomb potential  to
create BSEC. 


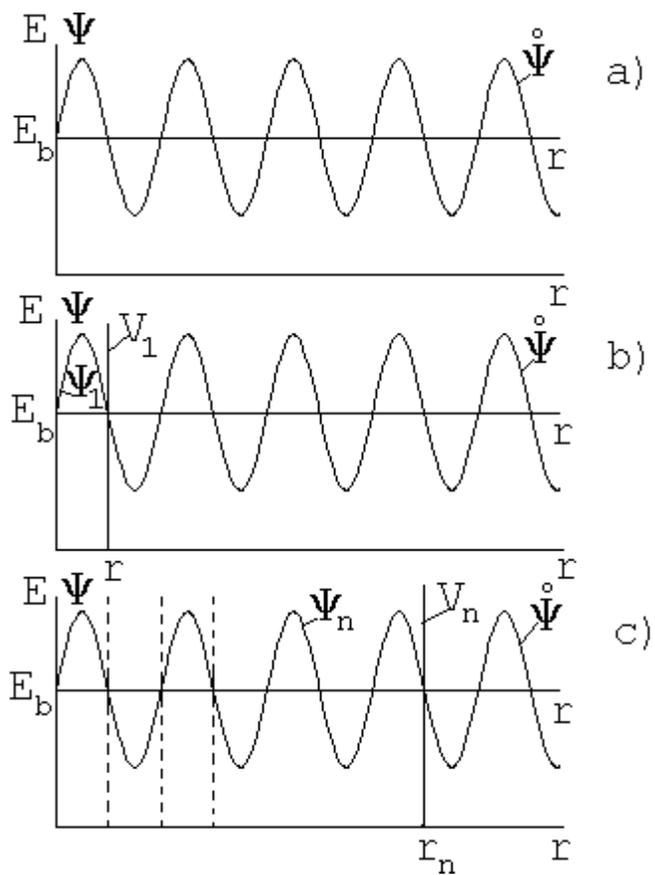
\end{document}